\documentclass[11pt,a4paper]{article}
\usepackage{jheppub}
\usepackage{extarrows}

\def\a{\alpha}
\def\b{\beta}
\def\g{\gamma}
\def\d{\delta}

\def\l{\lambda}

\def\t{\tau}

\def\ome{\omega}

\def\G{\Gamma}

\def\O{\Omega}

\def\RR{\mathbb R}

\def\ZZ{\mathbb Z}

\def\tad{N_{\rm{flux}}}

\def\o{\over}
\def\pa{\partial}

\def\td{\tilde}

\def\lp{\left(}
\def\rp{\right)}
\def\ls{\left[}
\def\rs{\right]}
\def\ie{{\it i.e. }}
\def\etc{{\it etc. }}

\preprint{MI-TH-813}

\title{On Fluxes in the $1^9$ Landau-Ginzburg Model}
\author[]{Katrin Becker,}
\author[]{Nathan Brady,}
\author[]{and Anindya Sengupta}
\affiliation[]{Mitchell Institute for Fundamental Physics and Astronomy,\\
Texas A\&M University College Station, TX 77843, USA}

\emailAdd{kbecker@physics.tamu.edu}
\emailAdd{bradyns@tamu.edu}
\emailAdd{anindya.sengupta@tamu.edu}
\abstract{In this paper we present a large class of flux backgrounds and solve the shortest vector problem in type IIB string theory on an orientifold of the  $1^9$ Landau-Ginzburg model.  }
\arxivnumber{2310.00770}
\begin{document}
\maketitle


\section{Introduction}

One of the most important, and challenging, questions in string theory is the existence and stability of vacua that may describe semi-realistic physics in four dimensions. The choice of internal manifold in string theory compactifications dictates many aspects of the four dimensional physics. In the case of the heterotic string, it was shown in \cite{Candelas:1985en} that under such reasonable assumptions that 
\begin{itemize}
    \item[a)] the vacuum be of the form $\mathcal{M}_4 \times M$ where $\mathcal M_4$ is a maximally symmetric four dimensional spacetime manifold and $M$ is a compact six-dimensional internal manifold, and
    \item[b)] there be unbroken $N=1$ supersymmetry in four dimensions,
\end{itemize}
$M$ is forced to be a Calabi-Yau three-fold, $\mathcal M_4$ is forced to be Minkowski, and the NS flux is not allowed to have a vacuum expectation value (vev). It also became immediately clear that there is no unique choice of such a vacuum configuration -- the moduli fields describing deformations of the internal manifold could not be given a set of unique values. Soon after the discovery of D-branes \cite{Polchinski:1995mt}, new supersymmetric vacua of type II string theories were found in \cite{Polchinski:1995sm}, with non-zero vev for the RR fluxes. Fluxes turn out to be good for multiple purposes. Naively, a Calabi-Yau compactification of the kind described above preserves $N=2$ supersymmetry in four dimensions. Incorporating fluxes provides a way \cite{Taylor:1999ii} to partially break supersymmetry from $N=2$ to $N=1$. It also generates a classical superpotential \cite{Taylor:1999ii, Giddings:2001yu} for moduli, raising the possibility of stabilizing some (or all) of them at a stable minimum of the potential. It was claimed to be possible to stabilize all complex structure moduli of Calabi-Yau manifolds in flux compactifications of type IIB or F-theory \cite{Giddings:2001yu, Giryavets:2003vd, Denef:2004dm, Denef:2005mm, Collinucci:2008pf}. However, it was conjectured recently in \cite{Bena:2020xrh, Bena:2021wyr} that, in models with a large number of complex structure moduli, the contribution of the flux to the D3-brane tadpole grows linearly with the number of stabilized moduli, a statement known as the tadpole conjecture. In such scenarios the price to pay for full moduli stabilization may be a violation of the tadpole cancellation condition. 

We will study in this paper some aspects of these compactification-related issues in type IIB string theory. Specifically, we will focus on a non-geometric compactification using an orientifold of the $1^9$ Landau-Ginzburg (henceforth LG) model orbifolded by a $\ZZ_3$ symmetry. The $1^9$ LG model is a tensor product of nine $N=2$ minimal models, each with level $k_i =1$, making a total central charge of $c=3$. It has world-sheet superpotential
\begin{equation}
\label{E:1^9-wsW}
    \mathcal W = \sum_{i=1}^9 x_i^3.
\end{equation}
In geometric compactifications, there is at least one Kähler modulus -- the overall size of the internal manifold. In general, therefore, one must be concerned with stabilizing both complex structure moduli and Kähler moduli. The fluxes generate a superpotential for the complex structure moduli, but the potential for the Kähler moduli is typically generated through non-perturbative effects. In order to avoid Kähler moduli altogether, and (try to) stabilize complex structure moduli by fluxes alone, we can look for compactifications with internal manifolds having $h^{1,1} =0$. String theory provides such examples where the internal manifolds are mirror duals to rigid\footnote{Calabi-Yau manifolds whose complex structure cannot be deformed.} Calabi-Yau manifolds. Since mirror symmetry interchanges complex and Kähler structures, these manifolds do not have Kähler moduli, and cannot be given a geometric interpretation. Nevertheless, they have a field theory description in terms of LG models. In a nutshell, this is the motivation to study such non-geometric compactifications. This idea was first pursued in \cite{Becker:2006ks} where supersymmetric flux backgrounds were found in the $1^9$ and $2^6$ LG models, leading to four dimensional Minkowski and Anti-de-Sitter spacetimes. Fluxes are described in these models using a combination of techniques from the world-sheet theory and the effective 4D theory. It was also argued in \cite{Becker:2006ks} that the flux superpotential is given by the standard GVW \cite{Gukov:1999ya} formula, 
\begin{equation}
\label{E:W-intro}
    W = \int_M G \wedge \O~,
\end{equation}
and that it receives no perturbative or non-perturbative correction thanks to a theorem concerning non-renormalization of the BPS tension of a D5-brane domain wall. It was then claimed in \cite{Becker:2006ks} that all complex structure moduli are stabilized via this flux-induced superpotential. A recent investigation of this claim in \cite{Becker:2022hse} revealed (also see \cite{Bardzell:2022jfh}) that not all moduli fields get a mass in the solutions presented in \cite{Becker:2006ks}. This does not rule out the possibility that some of the massless moduli are stable. The dependence of $W$ on moduli is given by (\ref{E:W-intro}) through how the holomorphic three-form $\O$ depends on them. One can compute an order-by-order expansion of $W$ (see Section 3, eqn. (\ref{E:holo-3-form-integral})) in the moduli deformation parameters, and some or all the massless moduli may be stabilized by terms at order higher than two. Thus, a systematic analysis of the supersymmetric vacua is necessary -- computing the number of massive moduli in each, and also the number of massless moduli stabilized at higher order -- to definitively understand the issue of moduli stabilization in these models. In the course of this exercise, the tadpole conjecture of \cite{Bena:2020xrh} can also be tested explicitly for these non-geometric compactification models. With this broad goal in mind, we launch a systematic search for Minkowski solutions in the $1^9/\ZZ_3$ model in this work.

Another interesting aspect is the recent classification \cite{Andriot:2022way, Andriot:2022yyj} of compactifications of type IIA/B supergravities down to 4D Minkowski, de Sitter, and anti-de Sitter spacetimes where the internal space is a 6D group manifold. The authors of these papers classify previously known solutions based on the $O_p /D_p$ sources present, and guided by this classification find new solutions in previously unexplored classes. Based on observation of a large number of solutions they propose some interesting conjectures, one of which is the {\it Massless Minkowski conjecture} stating that all Minkowski solutions of this kind must have at least one massless scalar field. Even though we study Minkowski solutions in a non-geometric compactification of type IIB string theory, we find that all solutions found in this model so far have massless fields. 

We begin by providing in Section \ref{S:Basic} the basic tools needed to compute all relevant quantities in the $1^9/\ZZ_3$ model. Conditions for type IIB compactifications to 4D Minkowski $N=1$ supersymmetric vacua are stated in the geometric setting, and then translated into the LG language. Then in Section \ref{S:Sec3} we present a large set of solutions satisfying these conditions. Using an exhaustive search algorithm described in Section \ref{S:Shortest-vector}, we find that there are no solutions in this model with flux tadpole $\leq 7$. We also present in Section \ref{S:Sec3} a large set of 8-flux-solutions which have flux tadpole  8. For all the aforementioned solutions, we also present the rank of the Hessian of the superpotential which equals the number of massive moduli. We do not analyze stabilization of massless fields at higher order presently, but show a convenient way of calculating derivatives of $W$ that will enable a computer to compute these corrections quite fast.


\section{Basics}
\label{S:Basic}

The conditions for type IIB string theory compactified to 4D with unbroken $N=1$ supersymmetry in the presence of background flux have been described in the literature many times. We begin by stating these conditions, formulated for compactifications on a geometric space $M$, maybe an orientifold of a Calabi-Yau three-fold. However, in this paper we are interested in backgrounds not described in terms of geometry but in terms of conformal field theory, in particular the LG model $1^9/\ZZ_3$. The aforementioned conditions will then have to be translated into LG language, which we do in the subsections that follow.

There is a flux-induced superpotential in compactifications of type IIB. It is given as usual by \cite{Becker:2006ks, Gukov:1999ya, Dasgupta:1999ss}
\begin{equation}
\label{E:W}
	W = \int_M G \wedge \O
\end{equation} 
where $\O$ is the holomorphic $(3,0)$-form, $G$ is the complex three-form flux obtained by combining the three-forms in the R-R and NS-NS sectors of type IIB string theory:
\begin{equation}
\label{E:G-flux}
G=H_{RR}-\tau H_{NS} , 
\end{equation} 
and $\t$ is the axio-dilaton:
\begin{equation}
\label{E:axio-dilaton}
    \t = C_0+i e^{-\phi}~.
\end{equation}
Unbroken supersymmetry demands that 
\begin{equation}
\label{E:susyG}
G=H_{RR}-\tau H_{NS}\in H^{(2,1)}(M) \oplus H^{(0,3)} (M)~. 
\end{equation}
In this paper we will focus on Minkowski solutions for which the superpotential $W$ vanishes, further constraining $G$:
\begin{equation}
\label{E:susy+Mink-G}
    G_{\rm{Mink}} \in H^{(2,1)}(M)~.
\end{equation}
Secondly, the tadpole cancellation condition requires 
\begin{equation}
\label{E:tad-cancln}
    \int_M H_{RR}\wedge H_{NS} + N_{D3} = Q_3({\rm O \textendash plane}),  
\end{equation} 
where $Q_3({\rm O \textendash plane})$ is the D3-brane charge of the orientifold planes, and $N_{D3}$ is the number of D3-branes in the geometry. 
Third, the fluxes have to obey the Dirac quantization conditions
\begin{equation}
\label{E:Dirac-Q}
    \int_\G G = N - \t M~,~~ \textrm{where~~} N, M \in \ZZ~,
\end{equation}
for any three-cycle $\G \in H_3(M, \ZZ)$.

We will now write down analogues of conditions (\ref{E:susy+Mink-G}, \ref{E:tad-cancln}, \ref{E:Dirac-Q}) in the LG language. Our aim is to be self-contained with regard to all necessary tools for computations. Detailed derivations can be found in \cite{Becker:2006ks, Becker:2022hse} and references therein.


\subsection{Cohomology}

The harmonic three-forms in the $1^9$ LG model are labelled by nine integers, which we assemble into a vector $\vec \ell=(\ell^1,\dots, \ell^9)$, such that
\begin{equation}
    \Omega_{\vec \ell} \in H^{(p,q)}(M) ~~\textrm{with}~~ p+q=3, \qquad \ell^i=1,2, \qquad \sum_{i=1}^9 \ell^i = 0 ~~ \textrm{mod}~~3.
\end{equation}
These arise from tensoring RR sector ground states \cite{Vafa:1989xc} in the building block minimal model, denoted $|\ell \rangle $, $\ell = 1,2$. The harmonic three-forms are classified into the four types of $(p,q)$-forms, $p+q=3$, as follows: 
\begin{equation}
    \begin{tabular}{|c|c|c|c|c|}
        \hline
        $\sum_i \ell^i$&9&12&15&18 \\
        \hline
        $H^{(p,q)}$&$H^{(3,0)}$&$H^{(2,1)}$&$H^{(1,2)}$&$H^{(0,3)}$ \\
        \hline
    \end{tabular}
\end{equation}
Therefore, condition (\ref{E:susy+Mink-G}) in the LG language becomes
\begin{equation}
\label{E:susy+Mink-G--LG}
    G \in \textrm{span}~\{ \O_{\vec \ell} : \ell^i \in \{1,2\}~ \textrm{and}~ \sum_i \ell^i = 12 \}~,
\end{equation}
which means that the vectors $\vec \ell$ are composed of exactly three $2$'s and six $1$'s.
We will consider the orientifold that combines worldsheet parity with the operator denoted by $g_1$ in \cite{Becker:2006ks}:
\begin{equation}
    g_1~:~(x_1, x_2, x_3, \ldots, x_9) \mapsto - (x_2, x_1, \ldots, x_9)~.
\end{equation}
What this means for the flux $G = \sum_{\vec \ell} B_{\vec \ell} \O_{\vec \ell}$~ is that it should be symmetric upon interchanging the first two entries of all $\vec \ell$ labels. This constrains $\O_{(1,2,\ldots)}$ and $\O_{(2,1,\ldots)}$ to either be turned on with equal relative strength or be simultaneously turned off\footnote{The entries in the ``$\ldots$" of the two $\O$'s in this sentence are identical of course.}. For ease of reference, we will say that these are {\it fluxes in the orientifold directions}. The fluxes of the kinds $\O_{(1,1,\ldots)}$ and $\O_{(2,2,\ldots)}$ are then referred to as {\it fluxes in the non-orientifold directions}. This orientifolding makes the span in (\ref{E:susy+Mink-G--LG}) have $63$ independent fluxes. To save ink while describing solutions in section \ref{S:Sec3}, we index the labels as specified in Appendix \ref{S:AppendixA}. For example, 
\begin{equation}
    \O_{(1,1,1,1,1,1,2,2,2)}=\O_1. 
\end{equation} 
This notation is particularly useful for orientifold directions. For example, 
\begin{equation} 
    \O_{(1,2,1,1,1,1,1,2,2)}+\O_{(2,1,1,1,1,1,1,2,2)}=\O_{36}.
\end{equation}


\subsection{Tadpole cancellation} 

The Bianchi identity for the RR 5-form is 
\begin{equation}
dF_5=H_{RR} \wedge H_{NS} +  \rho, 
\end{equation} 
and in a space-time described by geometry it can be integrated over the internal space $M$ to give the tadpole cancellation condition (\ref{E:tad-cancln}), which we restate:
\begin{equation}
\int_M H_{RR}\wedge H_{NS}+N_{D3}=Q_3({\rm O\textendash plane}),  
\end{equation} 
The topological nature of this condition allows us to formulate its analogue in the LG language by considering models that can be connected with some geometry by continuously varying moduli. For the orientifold we are considering, one gets \cite{Becker:2006ks}
\begin{equation}
    Q_3({\rm O \textendash plane}) = 12~,
\end{equation}
and the tadpole cancellation condition takes the form
\begin{equation}
\label{E:tad-cancln--LG}
    \int_M H_{RR} \wedge H_{NS} = \frac 1{\t - \bar \t} \int_M G \wedge \bar G = 12 - N_{D3}~.
\end{equation}
Here, $\bar G$ is obtained from $G=\sum_{\vec \ell} B_{\vec \ell} ~ \Omega_{\vec \ell}$~ by\footnote{Notaion: $\vec 1 = (1,1,1,1,1,1,1,1,1)$, $\vec 2 = (2,2,2,2,2,2,2,2,2)$, $\vec 3 = (3,3,3,3,3,3,3,3,3)$ \etc }
\begin{equation}
    \bar G = \sum_{\vec \ell} B^*_{\vec \ell} ~ \O_{\vec 3 - \vec \ell}
\end{equation}
The left hand side of eqn. (\ref{E:tad-cancln--LG}) is the contribution of the flux to the tadpole,
\begin{equation}
\label{E:tad}
    \tad := \frac 1{\t - \bar \t} \int_M G \wedge \bar G~,
\end{equation}
and is seen to be bounded above by $12$ for physical solutions. It (and the superpotential $W$ in eqn. (\ref{E:W})) can be computed using the Riemann bilinear identity. We will show some of these computations explicitly after introducing a basis of three-cycles in the LG language.


\subsection{Homology and flux quantization}
\label{subS:Flux-Q}

\begin{figure}[!h]
    \centering
    \includegraphics[scale=0.30]{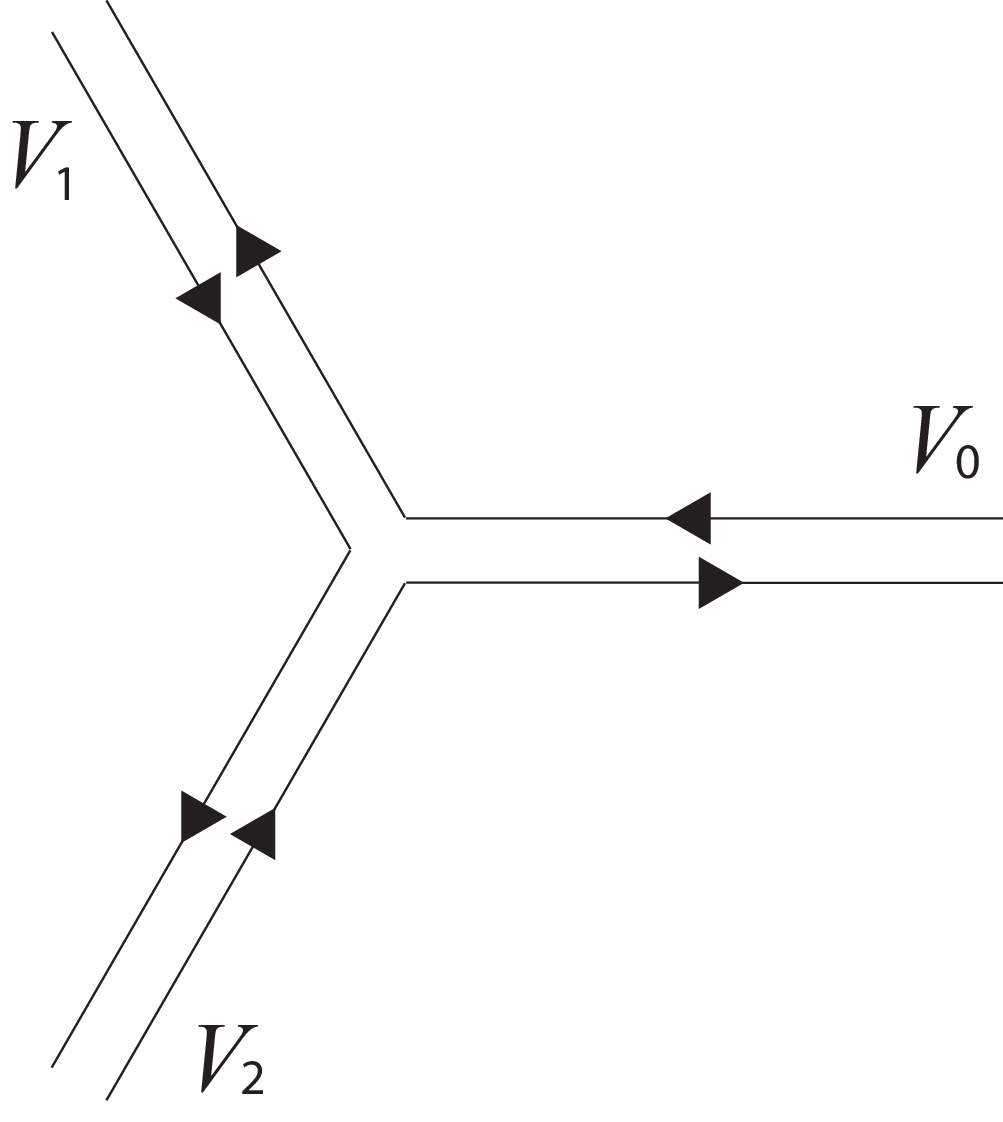}
    \caption{The ``pieces of cake": A-type D-branes in the LG model $x^3$~.}
    \label{Fig:cake}
\end{figure}

The $1^9$ LG model is a tensor product of nine copies of a minimal model with worldsheet superpotential $\mathcal W = x^3$. The A-type D-branes in this building block minimal model are described in the $\mathcal W$-plane by the positive real axis,
\begin{equation}
\label{E:A-Dbrane-W}
    \textrm{Im}~ \mathcal W = 0 ~,
\end{equation}
or, equivalently, in the $x$-plane as the contours $V_0, V_1, ~\textrm{and}~ V_2$ that look like the edges of three ``pieces of cake" (Figure \ref{Fig:cake}).
Clearly, they satisfy
\begin{equation}
    V_0 + V_1 + V_2 = 0~.
\end{equation}
A set of integral three-cycles for the $1^9/\ZZ_3$ model is built (see \cite{Becker:2006ks}) by tensoring nine $V_n$'s, and then $\ZZ_3$-completing them. Explicitly, these branes are
\begin{align}
\label{E:1^9/Z3-three-cycles}
    \G_{\vec n} &= \frac 1{\sqrt 3} \lp V_{\vec n} + V_{\vec n + \vec 1} + V_{\vec n + \vec 2} \rp = \frac 1{\sqrt 3} \lp \otimes_i V_{n_i} + \otimes_i V_{n_i + 1} + \otimes_i V_{n_i +2} \rp~,\\
    & \qquad \qquad \vec n = (n_1, \ldots, n_9), \qquad n_i = 0, 1, 2. \nonumber
\end{align}
$\ZZ_3$ acts on $\otimes_i V_{n_i}$ as a tensor product on each of the factors. On a factor $V_n$, it acts as $V_n \to V_{(n+1) ~\textrm{mod}~ 3}$. The set of cycles $\{\G_{\vec n}\}$ defined by (\ref{E:1^9/Z3-three-cycles}) is linearly dependent. It turns out that one can constrain $n_i$ to $n_i = 0,1,$ and further restrict $\vec n$'s to be the binary representations\footnote{written with nine binary digits, padding with zeroes on the left when necessary.} of the first $170$ non-negative integers to obtain an integral basis of three-cycles in the $1^9/\ZZ_3$ orbifold. Integrals of the fluxes through the three-cycles (see \cite{Becker:2006ks} for justification) are prescribed, with a normalization chosen for convenience, as follows. The pairing in the building block minimal model between the cycles $V_n$ and the RR sector ground states $|\ell \rangle$, $\ell = 1,2$, is given by
\begin{equation}
    \langle V | \ell \rangle = \tfrac{1}{ \left(\tfrac{-1 + \ome^\ell }{3} \right) \G\left(\tfrac{\ell}3\right) } \int_{V_n} x^{\ell -1} e^{- x^3} dx = \ome^{n \ell}~,
\end{equation}
where $\ome = e^{\tfrac{2 \pi i}{3}}$ is a cube root of unity. We are making the correspondence
\begin{equation}
    | \ell \rangle \xleftrightarrow[]{} \tfrac{1}{ \left(\tfrac{-1 + \ome^\ell }{3} \right) \G\left(\tfrac{\ell}3\right) } ~ x^{\ell -1}~.
\end{equation}
In the tensor product, this translates to
\begin{equation}
    \O_{\vec \ell} \xleftrightarrow[]{} | \vec \ell~ \rangle \xleftrightarrow[]{} \prod_{i = 1}^9  \tfrac{1}{ \left(\tfrac{-1 + \ome^{\ell^i} }{3} \right) \G\left(\tfrac{\ell^i}3\right) } ~ x_i^{\ell^i -1} ~,
\end{equation}
and
\begin{equation}
    \int_{V_{\vec n}} \O_{\vec \ell} = \prod_{i=1}^9  \int_{V_{n_i}} \frac{x_i^{\ell^i -1}}{\left(\tfrac{-1 + \ome^{\ell^i} }{3} \right) \G\left(\tfrac{\ell^i}3\right)} ~ e^{-x_i^3} ~ dx_i  = \ome^{\vec n \cdot \vec \ell}~.
\end{equation}

We are now ready to impose the flux quantization condition on the basis of three-cycles $\{\G_{\vec n}\}$, namely
\begin{equation}
\label{E:fluxQ}
    \int_{\Gamma_{\vec n}} G=N_{\vec n}-\tau M_{\vec n}~,  
\end{equation} 
where $N$ and $M$ are integers. This ensures flux quantization for any $\G \in H_3(M, \ZZ)$.
The result 
\begin{equation}
    \int_{\Gamma_{\vec n}} \Omega_{\vec \ell} = \sqrt{3} ~ \omega^{\vec n \cdot \vec \ell} 
\end{equation}
can be obtained by explicit computation and is very useful. 



\subsection{The homogenous basis of cycles}
\label{subS:W-basis}

At this point we would like to set up notation for a different basis of three-cycles, called the homogeneous basis, introduced in \cite{Becker:2006ks}. We will give its description in a pedestrian way, avoiding derivations, but highlighting how it makes certain computations convenient, resulting in simpler formulas. For the building block minimal model, let us define the cycles
\begin{subequations}
\label{E:W-cycles}
    \begin{align}
        W_0 &= V_0 + \ome V_1 + \ome^2 V_2 \\
        W_1 &= V_0 + \ome^2 V_1 + \ome V_2~.
    \end{align}
\end{subequations}
Their intersections are
\begin{equation}
    W_0 \cap W_1 = 0 = W_1 \cap W_0, \qquad W_1 \cap W_0 = 3 (1-\ome), \qquad W_0 \cap W_1 = 3(1-\ome^2)~.
\end{equation}
They have the following nice property:
\begin{subequations}
\label{E:int_W}
    \begin{align}
        \int_{W_0} x^n e^{-x^3} dx &= \d_{(n~ \textrm{mod}~3), 1}~(-1 + \ome^2) \G(\tfrac {n+1}3) \\
        \int_{W_1} x^n e^{-x^3} dx &= \d_{(n ~\textrm{mod}~3), 0} ~(-1 + \ome) \G(\tfrac {n+1}3) ~,
    \end{align}
\end{subequations}
resulting in the fact that each three-form flux $\O_{\vec \ell}$~ integrates to zero on all but one three-cycle obtained by tensoring nine $W_n$'s. Explicitly, let us denote by $C_{\vec \ell}$~ the cycles:
\begin{equation}
\label{E:C-cycles}
    C_{\vec \ell} := W_{\vec 2-\vec \ell} := \otimes_{i=1}^9 W_{2-\ell^i}~.
\end{equation}
The cycles $C_{\vec \ell}^*$ are given by
\begin{equation}
\label{E:C*-cycles}
    C^*_{\vec \ell} = W_{\vec \ell -\vec 1 } = \otimes_{i=1}^9 W_{\ell^i -1}~.
\end{equation}
For demonstration, $\vec \ell_1 = (1,1,1,1,1,1,2,2,2)$ $\Rightarrow$ $C_{\vec \ell_1} = W_{(1,1,1,1,1,1,0,0,0)} = W_1^{\otimes 6} \otimes W_0^{\otimes 3}$, and $C_{\vec \ell_1}^* = W_{(0,0,0,0,0,0,1,1,1)} = W_0^{\otimes 6} \otimes W_1^{\otimes 3}$. We then have
\begin{equation}
    \int_{C_{\vec \ell'}} \O_{\vec \ell} = 3^9 ~ \d_{\vec \ell , \vec \ell'}
\end{equation}
For each $\vec \ell$ such that $\O_{\vec \ell} \in H^{(2,1)}(M)$, we have
\begin{equation}
    C_{\vec \ell} \cap C^*_{\vec \ell}  = 3^9 (1- \ome^2)^3 (1- \ome)^6 = -i~ 3^{13}~ \sqrt{3}~.
\end{equation}
Computing the integrals to evaluate the superpotential (\ref{E:W}), or the flux tadpole (\ref{E:tad}) is much simpler if one employs the Riemann bilinear identity with a basis made of the $C$ cycles. For instance, 
\begin{equation}
\label{E:W-riem-bil}
    W = \int_M G \wedge \O = \sum_{C} \frac 1{C \cap C^*} \int_C G ~\int_{C^*} \O
\end{equation}
and, for each summand $B_{\vec \ell} ~ \O_{\vec \ell}$~ in $G = \sum_{\vec \ell} B_{\vec \ell} ~ \O_{\vec \ell}$~, only one cycle, namely $C_{\vec \ell}$~, contributes a non-zero value in the first integral on the right hand side of (\ref{E:W-riem-bil}).


\section{A large class of solutions}
\label{S:Sec3}

In this section we will present a large class of backgrounds and describe their properties. We will categorize solutions in terms of the number of $\Omega$'s turned on. We do so because of the following reason. It turns out that each non-zero component contributes at least 1 to the tadpole, implying that a lower bound for the flux tadpole\footnote{The orientifold we will consider has a tadpole value of $12$. Thus, the flux contribution to the tadpole can be maximally $12$. Therefore we need only turn on up to $12$ fluxes.} of a flux background with $n$ independent $\O_{\vec \ell}$ components turned on is $n$. Since one of the search criteria for flux backgrounds is the value of the flux tadpole, it makes sense to organize solutions in terms of its lower bound. For the cases when 1, 2, 3, or 4 components are turned on, we find that this lower bound is not saturated. We present for these cases the saturated lower bound of the flux tadpole, and all flux backgrounds that attain it.

As mentioned in (\ref{E:susy+Mink-G--LG}), the 63 independent harmonic $(2,1)$-form fluxes are labeled by vectors $\vec \ell$ composed of three $2$'s and six $1$'s. For convenience, we index them in this section  (also see Appendix \ref{S:AppendixA}) as follows: $I = (\a, A)$, with $\a \in \{1, \ldots , 35\} \cup \{57, \ldots , 63\}$ labeling $\vec \ell$'s whose first two entries are identical, and $A \in \{36, \ldots , 56\}$ labeling the ones of the form $(1,2, \ldots)$. We do not introduce an index for the $\vec \ell$'s of the form $(2,1, \ldots)$ since, as a result of orientifolding, turning on the flux $\O_{(1,2, \ldots)}$ would automatically turn on the flux $\O_{(2,1, \ldots)}$ with the same relative strength where the distribution of $1$'s and $2$'s in the two sets of ``$\ldots$" above are identical.  The generic flux background is a linear combination
\begin{equation}
\label{E:genericG}
	G =  \sum_{I=1}^{63} B_I \Omega_I 
\end{equation}
where the $G$-flux is as in (\ref{E:G-flux}). Here we have further simplified notation: $\O_{\vec l^I} = \O_I$. The coefficients $B_I$ are complex, so $126$ real numbers label each flux configuration.

How shall we proceed? We will be interested in solutions with $\t=\omega$. First, the flux quantization 
\begin{equation} \label{eq1}
\int_{\Gamma_{\vec n}} G=N_{\vec n}-\omega M_{\vec n} 
\end{equation} 
holds for any cycle in the basis $\{\G_{\vec n}\}$ of $170$ cycles. There are 170 $N$'s and 170 $M$'s, \ie in total 340 flux quantum numbers, which together with the real and imaginary parts of $B_I$ make a total of 466 real parameters. These parameters satisfy a total of $170 \times 2=340$ conditions which are the real and imaginary parts of eqn. (\ref{eq1}). We will then view 126 of the $N$'s and $M$'s as ``independent flux numbers'' and label them by $y_i$, $i=1,\dots,126$, and solve for $B_I$ in terms of the $y_i$. Collecting all real and imaginary parts of $B_I$ in a $126$-dimensional real vector $(b_i) = (\rm{Re} B_1, \rm{Im} B_1, \ldots, \rm{Re} B_{63}, \rm{Im} B_{63})$, this relationship reads $b_i = C_{ij} y_j$. The details of the matrix $C$ are not important in this section, but we bear in mind that flux quantization has been imposed in this way.

\subsection{Flux tadpole and massive moduli}

The two main properties of the solutions we will focus on are the flux tadpole $\tad$ (defined in (\ref{E:tad})), and the number of massive moduli fields.

\subsubsection{Flux tadpole}
The tadpole cancellation condition (\ref{E:tad-cancln--LG}), when $\t$ is  taken to be equal to $\ome$, becomes\footnote{by computing $\tad$ (\ref{E:tad}) by using the Riemann bilinear identity.} 
\begin{equation}
\label{E:tad-y's}
    \tad = 81\sum_I|B_I|^2 = \sum_{i, j} \mathcal Q_{ij} y_i y_j = 12 - N_{D3}, 
\end{equation} 
where $\mathcal Q$ is the symmetrized coefficient matrix of the homogeneous quadratic polynomial of $\{y_i : i=1(1)126\}$ obtained by substituting $b_i = C_{ij}y_j$ on the left hand side of (\ref{E:tad-y's}). Therefore, we should look for flux backgrounds with $\tad \leq 12$. By employing an exhaustive search algorithm, we verified that, in the orientifold of $1^9/\ZZ_3$ studied in this paper,
\begin{equation}
    \tad \geq 8~.
\end{equation}
Details of this result and the algorithm can be found in section \ref{S:Shortest-vector}. Thus, physical solutions in this model obey
\begin{equation}
    8 \leq \tad \leq 12~.
\end{equation}
One finds a large set of solutions in \cite{Becker:2006ks, Becker:2022hse}, some within this bound and some outside. We extend those results in this section in the following way. We first categorize solutions with respect to number of $\O_{\vec \ell}$'s turned on, find what the lowest value of $\tad$ can be for each category, and present all solutions attaining this greatest lower bound. We do this for up to 4-$\Omega$ solutions in subsection \ref{subS:solns}.

\subsubsection{Rank of the mass matrix}
Given a flux vacuum, an immediate question is whether this sits at a point in moduli space where all moduli are stabilized. If all scalar fields corresponding to deformations of the moduli around this point are massive, then no continuous deformation exists with zero energy cost, implying full moduli stabilization. However, all scalar fields being massive isn't a necessary condition. It is possible to have massless fields that are stabilized through interactions at higher order in deformation parameters. Here we focus on how many scalar fields are massive (and hence are stabilized at order two), and postpone the analysis of higher order deformations to future work.

The mass matrix of scalar fields in Minkowski solutions is given by a combination of the Hessian of the superpotential, and the inverse of the Kähler metric. It was shown\footnote{We also mention in passing that the mass matrix is positive semidefinite, ruling out tachyonic instabilities.} in \cite{Becker:2022hse} that, even though corrections to the Kähler potential are not under control, the rank of the physical mass matrix is the same as the rank of the Hessian of the superpotential $W$. Since the rank of the mass matrix is equal to the number of massive fields, and our goal is to count how many moduli are massive in a flux background, we will focus attention on computing the Hessian of $W$. Formulas for calculating the matrix elements of $\pa \pa W$ are given in \cite{Becker:2022hse} where the authors employ the Riemann bilinear identity using the basis $\{ \G_{\vec n} \}$ of cycles. We observe that using the homogeneous basis yields relatively simpler formulas, and significantly speeds up computations on a computer. This is especially useful for us since we analyze a large set of solutions.

The flux superpotential is given as usual by (\ref{E:W}):
\begin{equation}
	W = \int_M G \wedge \O
\end{equation}
in which the dependence of $W$ on all moduli comes from the holomorphic three-form $\O$ not to be confused with $\O_{\vec \ell}$. We use (\ref{E:W-riem-bil}), which we quote again for convenience:
\begin{equation}
\label{E:W-riem-bil-repeat}
    W = \int_M G \wedge \O = \sum_{C} \frac 1{C \cap C^*} \int_C G \int_{C^*} \O~, 
\end{equation}
with the cycles chosen from the homogeneous basis.
The second integral on the right hand side of eqn. (\ref{E:W-riem-bil-repeat}) encodes the full functional dependence of the superpotenial on deformations\footnote{We parameterize the deformations by local coordinates $ \{ t^{\vec \ell}, \vec \ell \in \{\vec \ell_I \} \}$. For convenience, let us write $t^I := t^{\vec \ell_I}$.}
of the moduli via the worldsheet superpotential
\begin{equation}
    \mathcal{W} (t^I) = \sum_{i=1}^9 x_i^3 - \sum_I t^I \vec x^{\vec \ell^I - \vec 1}~.
\end{equation}
For a generic flux background as in (\ref{E:genericG}), the superpotential evaluates to 
\begin{equation}
\label{E:W-W-basis1}
    \lp -i ~ 3^4 \sqrt 3 \rp W = \sum_\a B_\a \int_{C^*_\a} \O + \sum_A B_A \ls \int_{C^*_A} \O + \int_{C^{*'}_A} \O  \rs ~.
\end{equation}
Here,  we note that a flux $\O_A$ corresponding to the index $A$ is of the form $\O_{(12 \ldots)} + \O_{(21 \ldots)}$, which yields non-zero integrals on two distinct $C$-cycles instead of one -- the first summand is non-zero when integrated over $C_A$ as defined in (\ref{E:C-cycles}),  while the second summand gives non-zero integral over a $C$-cycle obtained from $C_A$ by interchanging its first two $W$-factors. It is this cycle which has been labeled temporarily as $C'_A$ in (\ref{E:W-W-basis1}). Now it remains to evaluate the integrals over $C^*$'s. We have, for an arbitrary cycle $\G$,
 \begin{equation}
 \label{E:holo-3-form-integral}
     \int_\G \O = \int_\G d^9 x~ \textrm{exp} \ls - \sum_{i=1}^9 x_i^3 + \sum_\a t^\a \lp \prod_{i=1}^9 x_i^{\ell_\a^i -1} \rp + \sum_A t^A (x_1 + x_2) \lp \prod_{i=3}^9 x_i^{\ell_A^i -1} \rp \rs
 \end{equation}
To compute Kähler covariant derivatives, we need the Kähler potential $K$. However, for Minkowski solutions, the following second Kähler covariant derivatives evaluated at the vacua are equal to the corresponding partial derivatives: $ D_{t^I} D_{t^J} W |_{t=0} = \pa_{t^I} \pa_{t^J} W |_{t=0}$~, $ D_\t D_{t^I} W |_{t=0} = \pa_\t \pa_{t^I} W |_{t=0}$~, where $D_{y} W = \pa_y W + (\pa_y K) W $, $y = (t^I, \t)$, and $D_\t D_\t W |_{t=0} = 0$. Combining all the ingredients provided above, it is straightforward to compute these second derivatives $\left[ \pa_{t^I} \pa_{t^J} W \right]|_{t = 0}$. We simply quote the results below:
\begin{subequations}
    \begin{align}
        k \frac{\pa^2}{\pa t^\a \pa t^\b} W |_{t=0} &= \sum_{\td \a} B_{\td \a} \prod_{i=1}^9 \d (\ell_{\td \a}^i + \ell_\a^i + \ell_\b^i, 4) \\
         k \frac{\pa^2}{\pa t^A \pa t^\b} W |_{t=0} &= \sum_{\td A} B_{\td A} ~.~ 2 ~ \d (\ell_\b^1 , 1) \d (\ell_\b^2 , 1) ~.~ \prod_{i=3}^9 \d (\ell_{\td A}^i + \ell_A^i + \ell_\b^i , 4) \\
         k \frac{\pa^2}{\pa t^A \pa t^B} W |_{t=0} &= \sum_{\td \a}B_{\td \a} ~.~ 2 ~ \d (\ell_{\td \a}^1, 1) \d (\ell_{\td \a}^2, 1) ~.~ \prod_{i=3}^9 \d (\ell_{\td \a}^i + \ell_A^i + \ell_B^i , 4)
    \end{align}
\end{subequations}
where 
\begin{equation}
    k = \frac{1}{\ls \G (\tfrac 13) \rs^3 \ls \G (\tfrac 23) \rs^6}~.
\end{equation}
Furthermore, the second derivatives of $W$ involving one or two derivatives with respect to the axio-dilaton are:
\begin{subequations}
    \begin{align}
        k' \frac{\pa^2}{\pa \t^2} W|_{t=0} &= 0 \\
        k' \frac{\pa^2}{\pa \t \pa t^\a} W|_{t=0} &= - \frac{1}{\t - \bar \t} \int_M \bar G \wedge \pa_{t^\a} \O ~=~ \frac i{\sqrt 3} B^*_\a \\
        k' \frac{\pa^2}{\pa \t \pa t^A} W|_{t=0} &= - \frac{1}{\t - \bar \t} \int_M \bar G \wedge \pa_{t^A} \O ~=~ \frac{2i}{\sqrt 3} B^*_A~,
    \end{align}
\end{subequations}
where
\begin{equation}
    k' = \frac{1}{\ls \G (\tfrac 13) \rs^6 \ls \G (\tfrac 23) \rs^3}~.
\end{equation}
This gives all matrix elements of the Hessian of the superpotential. Similar formulas can be derived for higher order derivatives to analyze stabilization of massless moduli at higher order.


\subsection{Solutions in terms of the number of $\Omega$'s} 
\label{subS:solns}

\subsubsection{1,2,3-$\Omega$ solutions} 

As a warm up let's discuss the simplest solutions, namely those in which only one, two, or three $\Omega$ components appear. These will not satisfy the tadpole cancellation condition. In what follows, we will sometimes refer to the flux tadpole $\tad$ as the tadpole for brevity.

\paragraph{1-$\Omega$ solutions:} 

First we consider the case where only one component in the non-orientifold direction is turned on, \ie 
\begin{equation}
\label{oneom}
    G=A \O_\a~.  
\end{equation} 
There is an $S_7$ symmetry which acts by interchanging the last 7 factors in the tensor product LG model. There is no $S_9$ symmetry since the first two factors are singled out by the action of the orientifold. Using this $S_7$ symmetry, we can take $\a=1$ or $\a=57$. The quantization condition in the first case becomes  
\begin{equation}
    \int_{\G_{\vec n}} G = A \int_{\G_{\vec n}}\Omega_{\vec \ell_1}=A \omega^{\vec n \cdot \vec \ell_1} \sqrt{3}=N_{\vec n}-\omega M_{\vec n}.
\end{equation} 
For this to hold for all $\G_{\vec n}$ in the integral basis, $A$ must be an integer multiple of $\frac 1{\sqrt{3}}$. The same argument applies to $\a = 57$. We find that the flux configuration that is properly quantized and attains the minimum value of tadpole is
\begin{equation}
    G={1\o \sqrt{3}} \O_\a~,  \qquad \a =1\quad {\rm or } \quad \a =57,  
\end{equation} 
and the minimal tadpole is 27. The quantization condition requires 
\begin{equation}
    \ome^{\vec n \cdot \vec \ell_\a}=N_{\vec n}-\omega M_{\vec n}~ .
\end{equation} 
Taking into account $1+\ome+\ome^2=0$ it is not difficult to see that it is always possible to choose flux numbers such that the above equation is satisfied for any $\vec n$. There are 16 massive scalars if $\a=1$, and 22 massive scalars if $\a=57$. Because of the $S_7 $ symmetry, any solution with $\vec \ell=\vec \ell_i$, $i=1,\dots, 35$ has tadpole 27 and lead to 16 massive scalars, and any solution with  $\vec \ell=\vec \ell_i$, $i=57,\dots, 63$ has tadpole 27 and leads to 22 massive scalars. 

In case a flux in an orientifold direction is involved, we find that the minimal tadpole value is attained by
\begin{equation}
    G={1\o \sqrt{3}} \Omega_{36} ~, 
\end{equation} 
where again the normalization is required by flux quantization. The minimal tadpole is twice the minimal tadpole of non-orientifold directions, 54, and there are 22 massive scalars. The $S_7$ symmetry then implies that the same results hold for any flux $\Omega_A$, with $A=36, \dots, 56$.

\paragraph{2-$\Omega$ solutions:} The smallest tadpole in this case is 18. The flux allowing this tadpole is of the form 
\begin{equation}
\label{E:2-om-solns1}
    G={i \o 3} \lp \O_1- \O_\a \rp , 
\end{equation} 
with $\a=2,\dots, 35,57,\dots,63$,  and a minimal tadpole of 18. The number of massive fields again depends on $\a$. For $\a=2,\dots,35,$ the number of massive fields can be 16, 24 or 26, while if $\a=57,\dots, 63,$ it can be 28 or 32. In this case, we have used the $S_7$ symmetry in taking the first term to be $\O_1$. 

Then there is the case in which we can take the first entry to be $\Omega_{57}$:
\begin{equation}
    G={i \o 3} \lp \O_{57}- \O_\a \rp , 
\end{equation} 
and without loss of generality\footnote{The choices $\a=1,\dots,35$ are covered in (\ref{E:2-om-solns1})} we can take $\a=58,\dots 63$. The number of massive fields is 22 for all $\a$ in this range. As in the 1-$\O$ case, the smallest tadpole is only achievable using non-orientifold directions.

We also note that any 2-$\O$ solution of the form 
\begin{equation}
\label{E:2-O-sol1}
    G={i \o 3} \left(\Omega_{\a_1}- \Omega_{\a_2}\right) , \quad \a_1, \a_2 \in \{1, \ldots , 35\} \cup \{57, \ldots , 63\}~,
\end{equation} 
is part of a more general set of solutions given by 
\begin{equation}
\label{E:2-O-sol2}
    G=\pm {i \o 3} \omega^p \left(\Omega_{\a_1}-\omega^q \Omega_{\a_2}\right) , 
\end{equation} 
where $p,q=0,1,2$ and the overall sign of $G$ and values of $p,q$ can be chosen independently for a total of 18 solutions for each choice of $\{\a_1, \a_2\}$. It is easy to see that if the flux (\ref{E:2-O-sol1}) is properly quantized so is (\ref{E:2-O-sol2}). Obviously this family of solutions has tadpole 18. The reason eq. (\ref{E:2-O-sol2}) is properly quantized is the elementary fact that there always exist integers $N$ and $M$ for which 
\begin{equation} \label{fact1}
{i\o \sqrt{3}}\left(\omega^{a}-\omega^{b}\right) =N-\omega M, 
\end{equation}
given any $a,b\in\ZZ$.

\paragraph{3-$\Omega$ solutions:} The smallest tadpole for a flux involving 3-$\Omega$'s is 27 and it is engendered by fluxes of the form
\begin{equation}\label{3oms1}
G={i \o 3} \left( \Omega_{1}+ \Omega_{\a}+\Omega_{\b}\right) , 
\end{equation} 
where $\a,\b$ can take any values $\a,\b=2,\dots,35, 57,\dots,63$
or 
\begin{equation}\label{3oms2}
G={i \o 3} \left( \Omega_{57}+ \Omega_{\a}+\Omega_{\b}\right) , 
\end{equation} 
with $\a,\b=58,\dots,63$. 

The number of massive fields does depend on $\a,\b$. 
If $\a,\b=2,\dots,35$ the number of massive fields takes one of the values in the set $\{16, 20, 24, 28, 22, 34, 29, 32, 30, 38, 42, 36, 40, 46\}$.
Again, also in this case there is a related set of properly quantized fluxes given by
\begin{equation}
G=\pm {i \o 3}\omega^p \left( \Omega_{1,57}+ \omega^q \Omega_{\a}+\omega^r \Omega_{\b}\right) , 
\end{equation} 
for $p,q,r\in \ZZ$. Evidently all of these solutions have tadpole $27$. Also in this case quantization is due to an elementary but not immediately obvious fact. Namely, there always exist integers $N$ and $M$ such that 
\begin{equation} \label{fact2}
{i\o \sqrt{3}}\left(
\omega^{a}
+\omega^{b}
+\omega^{c}\right) =N-\omega M, 
\end{equation}
for any $a,b,c\in\ZZ$.

\subsubsection{4-$\Omega$ solutions}

This is the first case in which the physical tadpole of 12 can be achieved with
\begin{equation} \label{4oms}
G={1\over 3\sqrt{3}} 
(-\Omega_{1}+
\Omega_{\a}+
\Omega_{\b}-
\Omega_{\g}) ~,
\end{equation} 
where the values for $(\a,\b,\g)$ can be found in the table below
\vskip .2cm
\footnotesize
\begin{tabular}{|c|c|c|c|c|c|c|}
\hline
(2,6,8)& (2,7,9)&(2,12,14)&(2,13,15)&(2,19,20)&(2,22,24)&(2,23,25)\hfill\\
(2,29,30)&(2,33,34)&(2,59,60)&(3,5,8)&(3,7,10)&(3,11,14)&(3,13,16)\\
(3,18,20)&(3,21,24)&(3,23,26)&(3,28,30)&(3,32,34)&(3,58,60)&(4,5,9)\\
(4,6,10)&(4,11,15)&(4,12,16)&(4,17,20)&(4,21,25)&(4,22,26)&(4,27,30)\\
(4,31,34)&(4,57,60)&(5,12,17)&(5,13,18)&(5,16,20)&(5,22,27)&(5,23,28)\\
(5,26,30)&(5,33,35)&(5,59,61)&(6,11,17)&(6,13,19)&(6,15,20)&(6,21,27)\\
(6,23,29)&(6,25,30)&(6,32,35)&(6,58,61)&(7,11,18)&(7,12,19)&(7,14,20)\\
(7,21,28)&(7,22,29)&(7,24,30)&(7,31,35)&(7,57,61)&(8,13,20)&(8,23,30)\\
(9,12,20)&(9,22,30)&(10,11,20)&(10,21,30)&(11,22,31)&(11,23,32)&(11,26,34)\\
(11,29,35)&(11,59,62)&(12,21,31)&(12,23,33)&(12,25,34)&(12,28,35)&(12,58,62)\\
(13,21,32)&(13,22,33)&(13,24,34)&(13,27,35)&(13,57,62)&(14,23,34)&(15,22,34)\\
(16,21,34)&(17,23,35)&(18,22,35)&(19,21,35)&(21,59,63)&(22,58,63)&(23,57,63)\\
\hline
\end{tabular}
\normalsize
\vskip .2cm
We note that also in this case there is a related family of fluxes with the same tadpole, {\it i.e.} tadpole 12 and are explicitly given by 
\begin{equation} \label{4oms1} 
G=\pm \omega^r {1\over 3\sqrt{3}} 
(-\Omega_{1}+
\omega^p\Omega_{\a}+
\omega^{q-p}\Omega_{\b}-
\omega^q \Omega_{\g}), 
\end{equation} 
where $r,p,q$ are integers. It is easy to verify that if (\ref{4oms}) is properly quantized so is eqn. (\ref{4oms1}). To do this it is useful to take (\ref{fact1}) and (\ref{fact2}) into account. Consequently for each flux in (\ref{4oms}) there are 54 fluxes given by including different phases and overall signs. The total number of 4-$\Omega$ fluxes with tadpole 12 is therefore $84\times 54=4536$. These background fluxes stabilize 16, 22 or 26 moduli fields. 

That these 4-$\Omega$ solutions are properly quantized can also be understood from the following simple fact. Given any 4 integers, $a,b,c,d\in \ZZ$ there exist integers $N,M\in\ZZ$ such that 
\begin{equation}
{1\o 3}(-\omega^a+\omega^b+\omega^c-\omega^d)=N-\omega M, 
    \end{equation} 
if and only if 
\begin{equation}
(-a-d+b+c)\!\!\!\mod 3=0.
\end{equation}
In particular, applied to the 4-$\Omega$ fluxes\footnote{Note: we return to our previous index notation, $\vec \ell=(\ell^1,\dots, \ell^9)$, introduced in section 2.} this means that any combination 
\begin{equation} \label{4oms2}
G={1\over 3\sqrt{3}} 
(-\Omega_{\vec \ell_a}+
\Omega_{\vec \ell_b}+
\Omega_{\vec \ell_c}-
\Omega_{\vec \ell_d}), 
\end{equation} 
will be properly quantized as long as 
\begin{equation} 
\vec n \cdot (-\vec \ell_a+\vec \ell_b+\vec \ell_c-\vec \ell_d)\!\!\!\mod 3=0, \qquad \forall \vec n. 
\end{equation}
A $9$ component vector $\vec w$, satisfying the condition $\vec n \cdot \vec w\mod 3=0$ for all $\vec n$, is a vector $\vec w$ whose entries are multiples of 3. It is not difficult to see that it is not possible to get any non-zero multiples of $3$ given any combination $-\vec \ell_a+\vec \ell_b+\vec \ell_c-\vec \ell_d$ since the components of the $\ell$'s are 1 or 2.  
The only solution is 
\begin{equation}
 -\vec \ell_a+\vec \ell_b+\vec \ell_c-\vec \ell_d=0. 
 \end{equation}
 Taking $a=1$
 \begin{equation}
 -\vec \ell_1+\vec \ell_b+\vec \ell_c-\vec \ell_d=0. 
 \end{equation}
 The solutions are exactly those quoted in the previous table.

\subsubsection{8-$\Omega$ solutions}

In this case the smallest tadpole is 8 and the corresponding fluxes take the form 
\begin{equation} \label{8oms}
G={1\over 9} 
(-\Omega_{1}+
\Omega_{\vec \ell_{a_2}}-
\Omega_{\vec \ell_{a_3}}+
\Omega_{\vec \ell_{a_4}}-
\Omega_{{\vec \ell_{a_5}}}+
\Omega_{\vec \ell_{a_6}}-
\Omega_{\vec \ell_{a_7}}+
\Omega_{\vec \ell_{a_8}}).
\end{equation} 
As in the 4-$\Omega$ case a necessary condition for the fluxes to be properly quantized is 
\begin{equation} 
(-\vec \ell_1+\vec \ell_{a_2}-\vec \ell_{a_3}+\vec \ell_{a_4}-
{\vec \ell_{a_5}}+\vec \ell_{a_6}-\vec \ell_{a_7}+\vec \ell_{a_8})\!\!\!\mod 3 =0,
\end{equation} 
but contrary to the 4-$\Omega$ case this condition is not sufficient. Aided by the computer it is possible to find those fluxes that turn out to be properly quantized. The table below gives the list of linearly independent solutions of this type by specifying $(a_2,\dots, a_8)$.
\vskip .4cm
\begin{center}
\footnotesize
\begin{tabular}{|c|c|c|c|}
\hline
(2,8,6,15,13,19,20)&(3,8,5,16,13,18,20)&(2,8,6,25,23,29,30)&(3,8,5,26,23,28,30)\\
(2,9,7,14,12,19,20)&(4,9,5,16,12,17,20)&(2,9,7,24,22,29,30)&(4,9,5,26,22,27,30)\\
(3,10,7,14,11,18,20)&(3,10,7,24,21,28,30)&(2,14,12,25,23,33,34)&(3,14,11,26,23,32,34)\\
(4,15,11,26,22,31,34)&(5,17,12,28,23,33,35)&&\\
\hline
\end{tabular}
\normalsize
\end{center}
All these flux backgrounds have $14$  massive fields.


\section{The shortest vector}
\label{S:Shortest-vector}

The shortest vector problem (SVP) looks for a non-zero vector with the smallest length in a lattice. The norm most commonly used to frame the question is the Euclidean norm, but the problem can be defined in a lattice with any norm. The quantity $\tad$, contribution of the fluxes to the tadpole, defines a norm in the lattice of quantized flux configurations, so finding a flux background with the minimum value of $\tad$ is an instance of the SVP. Algorithms to find the exact solution of SVP in an $n$-dimensional lattice are known, and follow one of three approaches: Lattice enumeration \cite{Kannan:1987}, Vornoi cell computation \cite{Micciancio:2013}, and Sieving \cite{Ajtai:2001}. All of these approaches have exponential or worse running time. There also exist polynomial time algorithms (based on basis reduction techniques) to solve the approximate version of SVP. Complexity-wise, it is known \cite{Ajtai:1998} that the SVP in $L_2$ norm is $\rm{NP}$-hard under randomized reductions. As far as we are aware, proving a similar hardness result under deterministic reductions is still an open problem. The approximate algorithms run faster, but only address the approximate version of SVP. We would like to ask the exact question instead: what is the smallest non-zero value of $\tad$ for flux vacua?

We adopt an exhaustive search algorithm combining sieving and enumeration to look for lattice vectors that are shorter than a fixed value. We describe this algorithm below, with the mathematica code implementing it available at \cite{GitHub}. The main result of this section is the following: there is no flux vacuum with $\tad \leq 7$. The minimum non-zero value of $\tad$ is 8, and is attained by a family of flux configurations.

Given a flux $G =  \sum_{I=1}^{63} B_I \Omega_I$ (\ref{E:genericG}), its contribution to $\tad$ in the Minkowski case is (\ref{E:tad})
\begin{equation}
    \tad = \frac{81 \sqrt 3}{2 ~ \rm{Im}\t} \sum_{I=1}^{63} |B_I|^2 \xlongequal{\t = \ome} 81 \sum_{I=1}^{63} |B_I|^2
\end{equation}
This is positive semidefinite, and zero if and only if $B_I = 0 ~\forall I$. It is most convenient to implement flux quantization (\ref{E:fluxQ}) on the integral basis of cycles $\G_{\vec n}$ as described in \cite{Becker:2006ks}. For convenience, it reads $\int_{\Gamma_{\vec n}} G=N_{\vec n}-\tau M_{\vec n}$. We separate the real and imaginary parts, $\vec b = (\rm{Re} B_1, \rm{Im} B_1, \ldots, \rm{Re} B_{63}, \rm{Im} B_{63})$,
and recast the flux quantization conditions in the form\footnote{The precise equations are supplied in \cite{GitHub}. The description of the algorithm only requires the form of the equation (\ref{E:b=cy}).}
\begin{equation}
\label{E:b=cy}
    b_i = C_{ij} y_j~, ~~~ i, j = 1(1)126~,
\end{equation}
where $y_i$ are some arrangement of the flux quantum numbers $N_{\vec n}, M_{\vec n}$, \ie $y_i \in \ZZ$, $i = 1(1)126$. This is a linearly independent system of equations. 

We observe two key facts. First, for each $I \in \{1, \ldots , 63\}$, $81 |B_I|^2$ is a homogeneous quadratic in the $y_i$'s with coefficients in $\ZZ$. Therefore, $N_{\rm{flux}}$ is non-negative integer-valued, and turning on $\O_I$ must contribute at least\footnote{This is the crudest lower bound for $\{81 |B_I|^2 : B_I \neq 0\}~.$} $1$ to $\tad$. This means that, if we want to find flux configurations with $\tad \leq T$, it suffices to consider $G = \sum_{I=1}^{63} B_I \Omega_I$ with $|\{I \in \{1, \ldots , 63\} : B_I \neq 0\}| \leq T$. Second, for each $I$, $81 |B_I|^2$ is a homogeneous quadratic polynomial in $y_i$'s with the symmetrized coefficient matrix positive definite. This latter fact plays a key role in sieving off lattice points in the second half of our method.

The first step in our algorithm is to turn off all but $T$ out of $63$ possible $B_I$'s. There are $63 \choose T$ ways\footnote{We can improve this by using the $S_7$ symmetry in the last seven factor CFT's in the model.} of doing it. For each choice $\{B_{i_1}, \ldots, B_{i_T}\}$, setting the remaining $B$'s to zero amounts to solving, over integers, a subsystem of $126 - 2T$ linear equations pulled from (\ref{E:b=cy}). Having solved this under-determined system, $\{B_{i_1}, \ldots, B_{i_T}\}$ are obtained as linear combinations of $2T$ arbitrary integers, say $c_i, i=1, \ldots , 2T$, in terms of which $\tad$ is expressed as
\begin{equation}
\label{E:Nflux-red}
    \tad^{\rm{red}} := \tad|_{B_I \neq 0 \Rightarrow I \in \{i_1, \ldots , i_T\}} = Q_{ij} c_i c_j~.
\end{equation}
The superscript ``red" stands for reduced, denoting the fact that we have reduced the number of independent integers. Clearly, the coefficients in $\tad^{\rm{red}}$ are also in $\ZZ$, and $\tad^{\rm{red}} \geq 0$, with $``="$ iff $c_i = 0 ~ \forall i$.

The second part of our algorithm is to check whether $\tad^{\rm{red}}$ attains non-zero values smaller or equal to $T$ for some choice of integers $c_i$, \ie we want to see if the level set $L_T = \{(c_1, \ldots , c_{2T}) : \tad^{\rm{red}} (\vec c) = T\} \subset \RR^{2T}$ has any integer points in it or in its interior. The level set is an ellipsoid since the symmetrized coefficient matrix $Q$ in (\ref{E:Nflux-red}) is positive definite\footnote{It follows from the positive-definiteness of the coefficient matrices for each $|B_I|^2$~.}. Let the eigenvalues of $Q$ be $\{\l_1, \ldots , \l_{2T} \}$, $0 < \l_1 \leq \l_2 \leq \ldots \leq \l_{2T}$, and the corresponding normalized eigenvectors be $\{\vec v_1, \ldots , \vec v_{2T} \}$. The intersection points of axis (axes)\footnote{There may be multiple if the lowest eigenvalue is degenerate. In case of degeneracies in any eigenvalue, the choices of the $\vec v_i$'s become ambiguous. Such a scenario hasn't occurred in practice, and is unimportant for the rest of the discussion.} along $\vec v_i$ corresponding to the lowest eigenvalue(s) with the ellipsoid are (among the) points on the ellipsoid that are farthest (in Euclidean norm) from the origin. Let us define the hypercube $\mathfrak C := \{\vec x \in \RR^{2T} : |x_i| \leq \sqrt{\frac T{\l_1}},  i=1(1)2T\}$. At all integer points outside $\mathfrak C$, \ie at points in $\ZZ^{2T} \cap \mathfrak C^c$, $\tad^{\rm{red}} > T$. So it is sufficient to evaluate $\tad^{\rm{red}}$ at all points in $\ZZ^{2T} \cap \mathfrak C$. Moreover, any point $\vec p$ in this set is in the exterior of $L_T$ if at least one of the following is satisfied:
\begin{equation}
\label{E:sievingEqns}
    | \vec p . \vec v_i | > \sqrt{\frac T{\l_i}}, ~~ i=1(1)2T~. 
\end{equation}
Using these criteria we sieve off points where evaluation of $\tad^{\rm{red}}$ is not necessary. At all remaining points in $\ZZ^{2T} \cap \mathfrak C$, we can evaluate $\tad^{\rm{red}}$ to check if values smaller or equal to $2T$ are attained. We call this algorithm the {\it Eigensieve} algorithm. Already in \cite{Becker:2006ks} solutions were known with fluxes contributing a value of $8$ to the tadpole. We set $T=7$ in our algorithm above to explicitly check that {\it there exists no solution with $\tad^{\rm{red}} \leq 7$, making $8$ the lowest value of  $\tad$ in the $1^9/\ZZ_3$ model.}

In summary, the Eigensieve algorithm rules out $\tad \leq 7$ as follows. First, it uses the observation that each non-zero flux contributes at least $1$ to $\tad$, thus dividing the problem into two sub-problems:
\begin{itemize}
    \item[a)] considering all possible ways of turning off all but $7$ fluxes;
    \item[b)] for each of the above, check whether $\tad^{\rm{red}} \leq 7$ is possible.
\end{itemize}
For the second part, a finite region in the lattice using the lowest eigenvalue $\l_1$ of $Q$, the coefficient matrix of $\tad^{\rm{red}}$, is carved out. Then the rest of the eigenvalues of $Q$ are used to sieve off more lattice points where evaluation is not necessary. The sieving conditions are (\ref{E:sievingEqns}). Then an explicit evaluation of $\tad^{\rm{red}}$ is done in the remaining lattice points.


\section{Conclusion}

The program of using Landau-Ginzburg models to describe flux vacua of type IIB compactifications was initiated in \cite{Becker:2006ks} with the goal that these would provide string vacua with all moduli fields stabilized. The underlying compactification manifolds before turning on fluxes are non-geometric since they are mirror duals to rigid 
Calabi-Yau manifolds, and therefore have no Kähler moduli. However, their world-sheet description is well understood in terms of Landau-Ginzburg models which at particular points in moduli space are equivalent to some Gepner models. Descriptions of geometric notions of forms, cycles, D-branes, orientifolds \etc in these models were developed from the world-sheet in \cite{Hori:2000ck, Recknagel:1997sb, Brunner:1999jq, Brunner:2003zm, Brunner:2004zd}. Reference \cite{Becker:2006ks} showed how to describe fluxes in this setting and presented explicit examples of flux vacua solutions that putatively stabilize all moduli.

More recently in \cite{Becker:2022hse} it was shown that all Minkowski vacua presented in \cite{Becker:2006ks} have a number of massless fields. A larger class of vacua was presented in the same paper, all of which have a large number of massless fields. Expanding the superpotential to higher-order terms may stabilize more (or all) moduli. To the best of our knowledge such a scenario has not been realized in any concrete example thus far. This prompts the need for a systematic search for solutions and investigation of their properties such as the number of massive fields, stabilization of massless fields by higher order terms in the superpotential, \etc In this paper we have taken a first step in this direction.

The key results of this work are as follows: 
\begin{itemize}
    \item[a)] A systematic search of solutions with the lowest value of $N_{\rm{flux}}$, organized by number of non-zero components, has been launched.  We present all solutions up to four components turned on, and a large set of solutions with eight components that saturate the minimum value of flux tadpole.
    \item[b)] The shortest vector problem for the $1^9/\ZZ_3$ model has been solved using an exact algorithm we call Eigensieve.
    \item[c)] We observe that the homogeneous basis of cycles can be used to simplify the formulas of derivatives of the superpotential with respect to moduli. We present these formulas for the second derivatives, which compute mass matrix elements. They increase computation speed significantly.
\end{itemize}

We are working on extending these results in a number of obvious ways:
\begin{itemize}
    \item[a)] The systematic search for solutions can be extended by increasing the number of non-zero components. The flux configurations known to satisfy $N_{\rm{flux}}=8$ are all $8$-$\O$ solutions. We have presented a large class of these in section \ref{S:Sec3}. The upper bound of $\tad$ in the $1^9 / \ZZ_3$ model, dictated by the tadpole cancellation condition, is $12$. A classification of all solutions characterized by $8 \leq \tad \leq 12$, along with the ranks of their mass matrices, will give a starting point for studying higher order corrections systematically. Some flux vacua with $\tad =12$ are known, but we do not yet have an exhaustive set of solutions with $\tad = 9, 10, 11, 12$.
    \item[b)]  Systematically computing mass matrices and their ranks to solutions with $8 \leq \tad \leq 12$ is computationally very expensive for Mathematica, even with the aid of parallel computations on a cluster. We think that it would be necessary to move away from symbolic computation in Mathematica to be able to achieve this task. Work is ongoing to make this process entirely numerical, and maybe use a lower level language/GPU's to speed up computations.
    \item[c)] We have not analyzed higher order terms in the superpotential in this work, leaving it to a forthcoming publication. We just mention that expanding the superpotential to higher orders is also made convenient by using the homogeneous basis.
    \item[d)] Finally, we aim to extend all our analyses to other Gepner models. 
\end{itemize}

\section*{Acknowledgements}
We would like to thank Timm Wrase, Muthusamy Rajaguru and Johannes Walcher for helpful discussions. Portions of this research were conducted with the advanced computing resources provided by Texas A\&M High Performance Research Computing. In particular we would like to thank Grigory Rogachev, Wilson Waldrop, Lisa Perez and Marinus Pennings for their invaluable help with setting up the computational part of this project. AS thanks William Linch for valuable feedback on an initial draft of the paper. NB and AS thank the Cynthia and George Mitchell Foundation for their hospitality during the Cook's Branch Workshop 2023 where part of this work was done. This work was partially supported by the NSF grant PHY-2112859.


\newpage 
\appendix
\section{A convenient indexing of the fluxes $\O_{\vec \ell}$}
\label{S:AppendixA}
For convenience we index in the following way the $(2,1)$-form fluxes of the $1^9/\ZZ_3$ model invariant under the orientifold action which exchanges the first two entries of the labels $\vec \ell$. We split them in three sets: non-orientifold fluxes labelled by $\vec \ell \sim (1,1, \ldots)$, orientifold fluxes, and non-orientifold fluxes labelled by $\vec \ell \sim (2,2,\ldots)$. Dropping commas for compactness, and denoting the index of a flux as a subscript,
\footnotesize
\begin{equation}\nonumber
\begin{split}
&\lp \O_{(111111222)}\rp_1, \lp \O_{(111112122)}\rp_2, \lp \O_{(111112212)}\rp_3, \lp \O_{(111112221)}\rp_4, \lp \O_{(111121122)}\rp_5, \lp \O_{(111121212)}\rp_6, \\
&\lp\O_{(111121221)}\rp_7, \lp \O_{(111122112)}\rp_8, \lp \O_{(111122121)}\rp_9, \lp \O_{(111122211)}  \rp_{10}, \lp \O_{(111211122)}\rp_{11}, \lp \O_{(111211212)}\rp_{12}, \\
&\lp \O_{(111211221)}\rp_{13}, \lp \O_{(111212112)}\rp_{14}, \lp \O_{(111212121)}\rp_{15}, \lp \O_{(111212211)}\rp_{16}, \lp \O_{(111221112)}\rp_{17},\lp \O_{(111221121)}\rp_{18}, \\
&\lp \O_{(111221211)}\rp_{19}, \lp \O_{(111222111)}\rp_{20}, \lp \O_{(112111122)}\rp_{21}, \lp \O_{(112111212)}\rp_{22}, \lp \O_{(112111221)}\rp_{23}, \lp \O_{(112112112)}\rp_{24}, \\
&\lp \O_{(112112121)}\rp_{25}, \lp \O_{(112112211)}\rp_{26}, \lp \O_{(112121112)}\rp_{27}, \lp \O_{(112121121)}\rp_{28}, \lp \O_{(112121211)}\rp_{29}, \lp \O_{(112122111)}\rp_{30}, \\
&\lp \O_{(112211112)}\rp_{31}, \lp \O_{(112211121)}\rp_{32}, \lp \O_{(112211211)}\rp_{33}, \lp \O_{(112212111)}\rp_{34},\lp \O_{(112221111)}\rp_{35}
\end{split} 
\end{equation} 
\begin{equation} \nonumber 
\begin{split} 
&\lp \O_{(121111122)}+\O_{(211111122)}) \rp_{36}, \lp \O_{(121111212)}+\O_{(211111212)} \rp_{37}, \lp \O_{(121111221)}+\O_{(211111221)} \rp_{38},\\
&\lp \O_{(121112112)}+\O_{(211112112)} \rp_{39},
\lp \O_{(121112121)}+\O_{(211112121)} \rp_{40}, \lp \O_{(121112211)}+\O_{(211112211)} \rp_{41},\\
&\lp \O_{(121121112)}+\O_{(211121112)} \rp_{42}, \lp \O_{(121121121)}+\O_{(211121121)} \rp_{43}, \lp \O_{(121121211)}+\O_{(211121211)} \rp_{44}\\
&\lp \O_{(121122111)}+\O_{(211122111)} \rp_{45}, \lp \O_{(121211112)}+\O_{(211211112)} \rp_{46}, \lp \O_{(121211121)}+\O_{(211211121)} \rp_{47},\\
&\lp \O_{(121211211)}+\O_{(211211211)} \rp_{48}, \lp \O_{(121212111)}+\O_{(211212111)} \rp_{49}, \lp \O_{(121221111)}+\O_{(211221111)} \rp_{50},\\
&\lp \O_{(122111112)}+ \O_{(212111112)} \rp_{51}, \lp \O_{(122111121)}+\O_{(212111121)} \rp_{52}, \lp \O_{(122111211)}+\O_{(212111211)} \rp_{53},\\
&\lp \O_{(122112111)}+\O_{(212112111)} \rp_{54}, \lp \O_{(122121111)}+\O_{(212121111)} \rp_{55}, \lp \O_{(122211111)}+\O_{(212211111)} \rp_{56}\\
\end{split} 
\end{equation} 
\begin{equation} \nonumber
\begin{split} 
&\lp \O_{(221111112)}\rp_{57}, \lp \O_{(221111121)}\rp_{58}, \lp \O_{(221111211)}\rp_{59}, \lp \O_{(221112111)}\rp_{60}, \lp \O_{(221121111)}\rp_{61}, \lp \O_{(221211111)}\rp_{62}, \\
&\lp \O_{(222111111)}\rp_{63}.
\end{split}
\end{equation}
\normalsize

\section{A basis of short fluxes}
\label{S:AppendixB}
In this appendix we present a basis of quantized fluxes that have small values of $\tad$. Let us first define the sets:
\footnotesize
\begin{subequations} \nonumber
\begin{align}
    \mathcal B_1 = \tfrac{i \omega}{ 9} \{ &   \left(\Omega _{12}-\Omega _{13}-\Omega _{17}+\Omega _{18}-\Omega _{22}+\Omega _{23}+\Omega _{27}-\Omega _{28}\right) , \\
    &   \left(\Omega _6-\Omega _7-\Omega _{17}+\Omega _{18}-\Omega _{22}+\Omega _{23}+\Omega _{31}-\Omega _{32}\right) , \\
    &  \left(\Omega _{11}-\Omega _{13}-\Omega _{17}+\Omega _{19}-\Omega _{21}+\Omega _{23}+\Omega _{27}-\Omega _{29}\right) , \\
    &   \left(\Omega _5-\Omega _7-\Omega _{17}+\Omega _{19}-\Omega _{21}+\Omega _{23}+\Omega _{31}-\Omega _{33}\right) , \\
    &  \left(\Omega _{12}-\Omega _{13}-\Omega _{14}+\Omega _{15}-\Omega _{22}+\Omega _{23}+\Omega _{24}-\Omega _{25}\right) , \\
    &   \left(\Omega _6-\Omega _7-\Omega _8+\Omega _9-\Omega _{22}+\Omega _{23}+\Omega _{24}-\Omega _{25}\right) , \\
    &  \left(\Omega _{11}-\Omega _{13}-\Omega _{14}+\Omega _{16}-\Omega _{21}+\Omega _{23}+\Omega _{24}-\Omega _{26}\right) , \\
    &   \left(\Omega _5-\Omega _7-\Omega _8+\Omega _{10}-\Omega _{21}+\Omega _{23}+\Omega _{24}-\Omega _{26}\right) , \\
    &   \left(\Omega _3-\Omega _4-\Omega _{14}+\Omega _{15}-\Omega _{22}+\Omega _{23}+\Omega _{31}-\Omega _{32}\right) , \\
    &   \left(\Omega _2-\Omega _4-\Omega _{14}+\Omega _{16}-\Omega _{21}+\Omega _{23}+\Omega _{31}-\Omega _{33}\right) , \\
    &   \left(\Omega _{11}-\Omega _{15}-\Omega _{17}+\Omega _{20}-\Omega _{21}+\Omega _{25}+\Omega _{27}-\Omega _{30}\right) , \\
    &   \left(\Omega _5-\Omega _9-\Omega _{17}+\Omega _{20}-\Omega _{21}+\Omega _{25}+\Omega _{31}-\Omega _{34}\right) , \\
    &   \left(\Omega _2-\Omega _9-\Omega _{14}+\Omega _{20}-\Omega _{21}+\Omega _{28}+\Omega _{31}-\Omega _{35}\right) , \\
    &   \left(\Omega _1-\Omega _4-\Omega _{12}+\Omega _{16}-\Omega _{21}+\Omega _{25}+\Omega _{31}-\Omega _{34}\right) \} ~,
\end{align}
\end{subequations}

\begin{subequations}\nonumber
\begin{align}
    \mathcal B_2 = \tfrac{1}{9} \{ & -i \left(\Omega _1-\Omega _4-\Omega _{21}-\Omega _{22}+\Omega _{25}+\Omega _{26}+\Omega _{51}-\Omega _{54}-\Omega _{57}+\Omega _{60}\right) , \\
    & -i \left(\Omega _1-\Omega _4-\Omega _{11}-\Omega _{12}+\Omega _{15}+\Omega _{16}+\Omega _{46}-\Omega _{49}-\Omega _{57}+\Omega _{60}\right) , \\
    & -i \left(\Omega _1-\Omega _4-\Omega _5-\Omega _6+\Omega _9+\Omega _{10}+\Omega _{42}-\Omega _{45}-\Omega _{57}+\Omega _{60}\right) , \\
    & -i \left(\Omega _1-\Omega _3-\Omega _{21}-\Omega _{23}+\Omega _{24}+\Omega _{26}+\Omega _{52}-\Omega _{54}-\Omega _{58}+\Omega _{60}\right) , \\
    & -i \left(\Omega _1-\Omega _3-\Omega _{11}-\Omega _{13}+\Omega _{14}+\Omega _{16}+\Omega _{47}-\Omega _{49}-\Omega _{58}+\Omega _{60}\right), \\
    & -i \left(\Omega _1-\Omega _3-\Omega _5-\Omega _7+\Omega _8+\Omega _{10}+\Omega _{43}-\Omega _{45}-\Omega _{58}+\Omega _{60}\right) , \\
    & -i \left(\Omega _1-\Omega _2-\Omega _{22}-\Omega _{23}+\Omega _{24}+\Omega _{25}+\Omega _{53}-\Omega _{54}-\Omega _{59}+\Omega _{60}\right) , \\
    & -i \left(\Omega _1-\Omega _2-\Omega _{12}-\Omega _{13}+\Omega _{14}+\Omega _{15}+\Omega _{48}-\Omega _{49}-\Omega _{59}+\Omega _{60}\right) , \\
    & -i \left(\Omega _1-\Omega _2-\Omega _6-\Omega _7+\Omega _8+\Omega _9+\Omega _{44}-\Omega _{45}-\Omega _{59}+\Omega _{60}\right) , \\
    & -i \left(\Omega _1-\Omega _7-\Omega _{21}-\Omega _{22}+\Omega _{28}+\Omega _{29}+\Omega _{51}-\Omega _{55}-\Omega _{57}+\Omega _{61}\right) , \\
    & -i \left(\Omega _1-\Omega _7-\Omega _{11}-\Omega _{12}+\Omega _{18}+\Omega _{19}+\Omega _{46}-\Omega _{50}-\Omega _{57}+\Omega _{61}\right) , \\
    & -i \left(\Omega _1-\Omega _{13}-\Omega _{21}-\Omega _{22}+\Omega _{32}+\Omega _{33}+\Omega _{51}-\Omega _{56}-\Omega _{57}+\Omega _{62}\right) , \\
    & -i \left(\Omega _1-\Omega _{11}-\Omega _{12}-\Omega _{23}+\Omega _{32}+\Omega _{33}+\Omega _{46}-\Omega _{56}-\Omega _{57}+\Omega _{63}\right) , \\
    & -i \left(\Omega _1-\Omega _5-\Omega _6-\Omega _{13}+\Omega _{18}+\Omega _{19}+\Omega _{42}-\Omega _{50}-\Omega _{57}+\Omega _{62}\right) , \\
    & -i \left(\Omega _1-\Omega _2-\Omega _3-\Omega _7+\Omega _9+\Omega _{10}+\Omega _{39}-\Omega _{45}-\Omega _{57}+\Omega _{61}\right) , \\
    & -i \left(\Omega _1-\Omega _2-\Omega _4-\Omega _6+\Omega _8+\Omega _{10}+\Omega _{40}-\Omega _{45}-\Omega _{58}+\Omega _{61}\right) , \\
    & -i \left(\Omega _1-\Omega _3-\Omega _4-\Omega _5+\Omega _8+\Omega _9+\Omega _{41}-\Omega _{45}-\Omega _{59}+\Omega _{61}\right) , \\
    & i \left(\Omega _1-\Omega _2+\Omega _3-\Omega _7+\Omega _9-\Omega _{10}-\Omega _{37}+\Omega _{44}+\Omega _{57}-\Omega _{61}\right) , \\
    & i \left(\Omega _1-\Omega _2+\Omega _4-\Omega _6+\Omega _8-\Omega _{10}-\Omega _{38}+\Omega _{44}+\Omega _{58}-\Omega _{61}\right) , \\
    & i \left(\Omega _1+\Omega _2-\Omega _3-\Omega _7-\Omega _9+\Omega _{10}-\Omega _{36}+\Omega _{43}+\Omega _{57}-\Omega _{61}\right)   \}~,
 \end{align}
 \end{subequations}
 \begin{subequations}\nonumber
 \begin{align}
    \mathcal B_3 = \tfrac{i \omega}{9} \{ &  \left(\omega  \Omega _4-\omega  \Omega _{25}-\omega  \Omega _{26}+\omega  \Omega _{54}-\omega  \Omega _{60}-\Omega _{13}+\Omega _{32}+\Omega _{33}-\Omega _{56}+\Omega _{62}\right) ,\\
    &   \left(\omega  \Omega _4-\omega  \Omega _{15}-\omega  \Omega _{16}+\omega  \Omega _{49}-\omega  \Omega _{60}-\Omega _{23}+\Omega _{32}+\Omega _{33}-\Omega _{56}+\Omega _{63}\right) , \\
    &   \left(\omega  \Omega _4-\omega  \Omega _9-\omega  \Omega _{10}+\omega  \Omega _{45}-\omega  \Omega _{60}-\Omega _{23}+\Omega _{28}+\Omega _{29}-\Omega _{55}+\Omega _{63}\right) , \\
    &   \left(\omega  \Omega _3-\omega  \Omega _{24}-\omega  \Omega _{26}+\omega  \Omega _{54}-\omega  \Omega _{60}-\Omega _{12}+\Omega _{31}+\Omega _{33}-\Omega _{56}+\Omega _{62}\right) , \\
    &   \left(\omega  \Omega _3-\omega  \Omega _{14}-\omega  \Omega _{16}+\omega  \Omega _{49}-\omega  \Omega _{60}-\Omega _{22}+\Omega _{31}+\Omega _{33}-\Omega _{56}+\Omega _{63}\right) , \\
    &   \left(\omega  \Omega _3-\omega  \Omega _8-\omega  \Omega _{10}+\omega  \Omega _{45}-\omega  \Omega _{60}-\Omega _{22}+\Omega _{27}+\Omega _{29}-\Omega _{55}+\Omega _{63}\right) , \\
    &   \left(\omega  \Omega _2-\omega  \Omega _{24}-\omega  \Omega _{25}+\omega  \Omega _{54}-\omega  \Omega _{60}-\Omega _{11}+\Omega _{31}+\Omega _{32}-\Omega _{56}+\Omega _{62}\right) , \\
    &   \left(\omega  \Omega _2-\omega  \Omega _{14}-\omega  \Omega _{15}+\omega  \Omega _{49}-\omega  \Omega _{60}-\Omega _{21}+\Omega _{31}+\Omega _{32}-\Omega _{56}+\Omega _{63}\right) , \\
    &  \left(\omega  \Omega _2-\omega  \Omega _8-\omega  \Omega _9+\omega  \Omega _{45}-\omega  \Omega _{60}-\Omega _{21}+\Omega _{27}+\Omega _{28}-\Omega _{55}+\Omega _{63}\right) , \\
    &   \left(\omega  \Omega _7-\omega  \Omega _9-\omega  \Omega _{10}+\omega  \Omega _{45}-\omega  \Omega _{61}-\Omega _{23}+\Omega _{25}+\Omega _{26}-\Omega _{54}+\Omega _{63}\right) , \\
    &   \left(\omega  \Omega _6-\omega  \Omega _8-\omega  \Omega _{10}+\omega  \Omega _{45}-\omega  \Omega _{61}-\Omega _{22}+\Omega _{24}+\Omega _{26}-\Omega _{54}+\Omega _{63}\right) , \\
    &   \left(\omega  \Omega _5-\omega  \Omega _8-\omega  \Omega _9+\omega  \Omega _{45}-\omega  \Omega _{61}-\Omega _{21}+\Omega _{24}+\Omega _{25}-\Omega _{54}+\Omega _{63}\right) , \\
    &   \left(\omega  \Omega _4-\omega  \Omega _{23}-\omega  \Omega _{26}+\omega  \Omega _{53}-\omega  \Omega _{59}-\Omega _{15}+\Omega _{32}+\Omega _{34}-\Omega _{56}+\Omega _{62}\right) , \\
    &   \left(\omega  \Omega _4-\omega  \Omega _{13}-\omega  \Omega _{16}+\omega  \Omega _{48}-\omega  \Omega _{59}-\Omega _{25}+\Omega _{32}+\Omega _{34}-\Omega _{56}+\Omega _{63}\right) , \\
    &   \left(\omega  \Omega _4-\omega  \Omega _7-\omega  \Omega _{10}+\omega  \Omega _{44}-\omega  \Omega _{59}-\Omega _{25}+\Omega _{28}+\Omega _{30}-\Omega _{55}+\Omega _{63}\right) , \\
    &   \left(\omega  \Omega _7-\omega  \Omega _{23}-\omega  \Omega _{29}+\omega  \Omega _{53}-\omega  \Omega _{59}-\Omega _{18}+\Omega _{32}+\Omega _{35}-\Omega _{56}+\Omega _{62}\right) , \\
    &   \left(\omega  \Omega _7-\omega  \Omega _{13}-\omega  \Omega _{19}+\omega  \Omega _{48}-\omega  \Omega _{59}-\Omega _{28}+\Omega _{32}+\Omega _{35}-\Omega _{56}+\Omega _{63}\right) , \\
    &   \left(\omega  \Omega _{13}-\omega  \Omega _{23}-\omega  \Omega _{33}+\omega  \Omega _{53}-\omega  \Omega _{59}-\Omega _{18}+\Omega _{28}+\Omega _{35}-\Omega _{55}+\Omega _{61}\right) , \\
    & -  \left(\omega  \Omega _7-\omega  \Omega _9+\omega  \Omega _{10}-\omega  \Omega _{44}+\omega  \Omega _{61}-\Omega _{23}+\Omega _{25}-\Omega _{26}+\Omega _{53}-\Omega _{63}\right) , \\
    & -  \left(\omega  \Omega _7+\omega  \Omega _9-\omega  \Omega _{10}-\omega  \Omega _{43}+\omega  \Omega _{61}-\Omega _{23}-\Omega _{25}+\Omega _{26}+\Omega _{52}-\Omega _{63}\right) , \\
    & -  \left(\omega  \Omega _6+\omega  \Omega _8-\omega  \Omega _{10}-\omega  \Omega _{42}+\omega  \Omega _{61}-\Omega _{22}-\Omega _{24}+\Omega _{26}+\Omega _{51}-\Omega _{63}\right) \} ~,
\end{align}
\end{subequations}

\begin{subequations} \nonumber
\begin{align}
    \mathcal B_4 = \tfrac{-i \omega}{9} \{ &  \left(2 \Omega _1-2 \Omega _2+\Omega _{12}-\Omega _{14}+\Omega _{23}-\Omega _{25}-\Omega _{33}+\Omega _{34}\right) , \\
    &   \left(2 \Omega _1-2 \Omega _3+\Omega _{11}-\Omega _{14}+\Omega _{23}-\Omega _{26}-\Omega _{32}+\Omega _{34}\right) , \\
    &   \left(2 \Omega _1-2 \Omega _4+\Omega _{11}-\Omega _{15}+\Omega _{22}-\Omega _{26}-\Omega _{31}+\Omega _{34}\right) , \\
    &   \left(2 \Omega _1-2 \Omega _5+\Omega _{12}-\Omega _{17}+\Omega _{23}-\Omega _{28}-\Omega _{33}+\Omega _{35}\right) , \\
    &   \left(2 \Omega _1+\Omega _6-2 \Omega _{11}-\Omega _{17}+\Omega _{23}-\Omega _{29}-\Omega _{32}+\Omega _{35}\right) , \\
    &  \left(2 \Omega _1+\Omega _6+\Omega _{13}-\Omega _{19}-2 \Omega _{21}-\Omega _{27}-\Omega _{32}+\Omega _{35}\right) \} ~,
\end{align}
\end{subequations}
\begin{subequations}\nonumber
\begin{align}
    \mathcal B_5 &= \{ - \tfrac{i \omega}{3}  \left(\Omega _1-\Omega _{57}\right) \}~,   \\
    \mathcal B_6 &= \{  - \tfrac i{9} \left(\omega  \Omega _1+2 \omega  \Omega _3-\omega  \Omega _7+\omega  \Omega _{10}+ (3+ 2 \omega ) \Omega _{21}-\omega  \Omega _{24}+\omega  \Omega _{28}-\omega  \Omega _{30} \right) \} ~,
\end{align}
\end{subequations}
\normalsize 
where we have taken the liberty to use the notation that a numerical ``overall factor" multiplies all elements of a set. One finds that 
\begin{equation} \nonumber
    \mathcal B = \lp \cup_{i=1}^6 \mathcal B_i \rp  \cup  \lp \cup_{i=1}^6 \mathcal \ome B_i \rp
\end{equation}
is a basis of flux vectors in this model. Any quantized flux is an integer linear combination of the fluxes in $\mathcal B$. All fluxes in $\mathcal B_i$ (and hence obviously in $\ome \mathcal B_i$) have flux tadpole values of $8, 12, 12, 14, 18, 17$ for $i = 1, 2, 3, 4, 5, 6$ respectively.

\newpage

\providecommand{\href}[2]{#2}\begingroup\raggedright
\endgroup

\end{document}